\documentclass[]{iopart}
\usepackage{epsfig}
\begin{document}

\title{$N$-qubit states as points on the Bloch sphere}

\author{H M\"akel\"a and A Messina}

\address{Dipartimento di Scienze Fisiche ed Astronomiche, Universit\`a di Palermo, via Archirafi 36, 90123 Palermo, Italy}
\ead{harmak@gmail.com}
\begin{abstract}
We show how the Majorana representation can be used to express the pure states of an $N$-qubit system as points on the Bloch sphere. We compare this geometrical representation of $N$-qubit states with an alternative one, proposed recently by the present authors.  
 \end{abstract}

\section{Introduction}
In 1932 E. Majorana introduced a way to represent the pure states of a spin-$S$ particle 
in terms of symmetrized states of $2S$ spin-$\frac{1}{2}$ particles \cite{Majorana32,Bloch45}. 
This representation, known as the Majorana representation, 
makes it possible to express spin-$S$ states geometrically as $2S$ points on the Bloch sphere. 
Remarkably, in this geometrical description a rotation of a spin-$S$ state corresponds 
to a rigid-body rotation of the corresponding points on the sphere.  
This property has made the Majorana representation a useful tool in many different fields, such as  
 in studying the symmetries of spinor Bose-Einstein condensates   
\cite{Barnett06,Barnett07,Makela07,Barnett09,Lamacraft09}, 
in the context of reference frame alignment \cite{Kolenderski08}, 
in helping to define so-called anticoherent states \cite{Zimba06}, and 
in the geometrical description of the states of an $n$-level system \cite{Bijurkar06}. 
A variant of the Majorana representation was used to study 
the orbits of $SO(3)$ in \cite{Bacry74}.

We presented recently a mapping that associates the pure states of an $N$-qubit system with a polynomial \cite{Makela09}. 
We showed that it provides a new way of detecting the separability of pure $N$-qubit states. 
In this article we describe how this method can be used to express pure $N$-qubit states geometrically. 
At the moment this description works only for pure states; a generalization to mixed states is not known. 
We compare this geometrical description with the Majorana one, concentrating on their behaviour in different types of rotations. 
We argue that our method is preferable when  $N$-qubit systems are studied. Our work on the geometrical representations 
of $N$-qubit states is motivated by the usefulness of this type of approach in other contexts \cite{Barnett06,Barnett07,Makela07,Barnett09,Lamacraft09,Kolenderski08,Zimba06}.

\section{The Majorana representation\label{Majo}}
In this section we give a short introduction to the Majorana 
representation; a more detailed discussion can be found in \cite{Bloch45}.  
Let $\{|S,M\rangle,\, M=-S,-S+1,\ldots, S\}$ be the basis for the spin states 
of a spin-$S$ particle. An arbitrary pure spin state can be written as 
\begin{equation} 
|\xi\rangle=\sum_{M=-S}^S\,\xi_M\,|S,M\rangle,  
\end{equation}
where $\xi_M$ is a complex number for each $M$. This state can be expressed in terms of $2S$ spin-$\frac{1}{2}$ particles by coupling their spins  
 in such a way that the total spin equals $S$. We denote the basis of the spin-$\frac{1}{2}$ particle labelled by $j$ by $\{|+\rangle_j^{\textrm{M}},|-\rangle_j^{\textrm{M}}\}$, where the superscript M refers to the Majorana representation. 
The (uncoupled) basis vectors of the $2S$ particle system have the form
\begin{equation}\label{tensorbasis}
|\gamma_1\gamma_2\cdots\gamma_{2S}\rangle\equiv 
|\gamma_1\rangle_1^{\textrm{M}}\otimes|\gamma_2\rangle_2^{\textrm{M}}\otimes\cdots\otimes |\gamma_{2S}\rangle_{2S}^{\textrm{M}},
\end{equation}
where each $\gamma_i$ is either $+$ or $-$.
We denote the spin operator of the $j$th spin-$\frac{1}{2}$ particle by $\mathbf{s}_j$, so  
the operator for the total spin is 
\begin{equation}
\mathbf{S}_{\textrm{tot}}\equiv\mathbf{s}_1+\mathbf{s}_2+\cdots +\mathbf{s}_{2S}.  
\end{equation} 
The Majorana representation uses the eigenstates of 
$\mathbf{S}_{\textrm{tot}}^2$ related to its largest eigenvalue $S(S+1)$.   
The eigenvector with maximal projection in the $z$-direction corresponds to the state $|S,S\rangle$:
\begin{equation}
|S,S\rangle= |\underbrace{++\cdots +}_{2S}\rangle. 
\end{equation}
The expressions for the rest of the spin-$S$ states $|S,M\rangle$ 
can be obtained from this by acting on it repeatedly with the lowering operator 
$\mathbf{S}_-=\mathbf{s}_{1-}+\mathbf{s}_{2-}+\cdots +\mathbf{s}_{2S-}$. 
The general result is \cite{Bloch45}
\begin{equation}\label{SM}
|S,M\rangle=\frac{1}{(2S)!}{2S \choose S+M}^{1/2}\!\!\!\sum_{P} 
|\underbrace{++\cdots +}_{S+M}\underbrace{--\cdots -}_{S-M}\rangle
\end{equation}
where $\sum_{P}$ denotes the sum over all permutations of the $\pm$ -signs. As we show next, 
the components of the spin-$S$ state $|\xi\rangle$ can be obtained using 
(\ref{SM}) if the constituent spin-$\frac{1}{2}$ states are known.
We define $2S$ spin-$\frac{1}{2}$ states  
\begin{equation}
\varphi_k\equiv\alpha_k |+\rangle_k^{\textrm{M}}+\beta_k |-\rangle_k^{\textrm{M}},\quad k=1,2,\ldots,2S,  
\end{equation}
and define a spin-$S$ state $|\tilde{\xi}\rangle$ as a symmetrized tensor product of these states
\begin{equation}\begin{array}{lll}
\label{xiprime}
|\tilde{\xi}\rangle &=& \displaystyle\frac{A}{(2S)!}\sum_{P}\bigotimes_{k=1}^{2S}\varphi_k. 
\end{array}\end{equation}   
Here $A$ is an arbitrary non-zero complex number. 
A straightforward calculation utilizing (\ref{SM}) gives  
\begin{equation}\begin{array}{lll}\label{xistate}
|\tilde{\xi}\rangle &=&\displaystyle\sum_{M=-S}^{S} \underbrace{A\,C_M {2S \choose S+M}^{-1/2}|}_{=\tilde{\xi}_M}S,M\rangle,
\end{array}\end{equation}
where
\begin{equation}\label{CM}
C_M= \frac{1}{(S+M)! (S-M)!}\sum_{P_ k} \prod_{ k=1}^{S+M}\alpha_{ k}\prod_{ k=S+M+1}^{2S}\beta_{ k} 
\end{equation}
and $\sum_{P_k}$ denotes the sum over all permutations of the indices $k$ of $|\pm\rangle_k$. Notice that $\sum_{P_k}$ and $\sum_{P}$ 
are different operations.
This result shows that when the states of the spin-$\frac{1}{2}$ particles are known, the components $\tilde{\xi}_M$ of the corresponding 
spin-$S$ state $|\tilde{\xi}\rangle$ can be obtained from equations  (\ref{xistate}) and (\ref{CM}). Note that changing 
the normalization and phase of the spin-$\frac{1}{2}$ vectors does not change  
the physical state, defined as $|\tilde{\xi}\rangle\langle\tilde{\xi}|/\langle\tilde{\xi}|\tilde{\xi}\rangle$.

In order to obtain the spin-$\frac{1}{2}$ vectors if $|\tilde{\xi}\rangle$ is known, we define the {\it Majorana polynomial} 
\begin{equation}
\label{MaPo}
P_{Majo}(|\tilde{\xi}\rangle;x) \equiv \sum_{M=-S}^{S}\!{2S \choose S+M}^{1/2}\!\!\!\!\!\tilde{\xi}_M\,x^{S+M}.
\end{equation}
A straightforward calculation utilizing   (\ref{xistate}) and (\ref{CM}) shows that $P_{Majo}(|\tilde{\xi}\rangle;x)$ can be written as  
\begin{equation}\label{MaPo2}
P_{Majo}(|\tilde{\xi}\rangle;x)=\displaystyle A\prod_{k=1}^{2S}(\alpha_k x+\beta_k),
\end{equation}
which means that the zeros of the Majorana polynomial are given by 
$x_k\equiv -\beta_k/\alpha_k$. With the help of these ratios it is possible to construct, up to a normalization and phase,  the 
spin-$\frac{1}{2}$ vectors $\varphi_k$. 

Using the roots of the Majorana polynomial, every state $|\xi\rangle\langle\xi|$ can be expressed in terms 
of $2S$ points on the Bloch sphere. This is obtained by writing the roots as 
\begin{equation}\label{angles}
\tan\frac{\theta_k}{2}\, \exp(\rmi\varphi_k)= x_k, 
\end{equation} 
which defines the locations of the $2S$ points, given using the angular variables  of spherical coordinates. 
Now $\alpha_k=0$ corresponds to $\theta_k=\pi$, which means that if the degree of (\ref{MaPo2}) is $2S-l$, then 
$l$ of the points are at the south pole. An important property of the Majorana representation  
is that each state $|\xi\rangle\langle\xi|$ determines a unique point configuration and vice versa.

It can be shown that if the Majorana polynomial is defined as in (\ref{MaPo}) a rotation of the vector $|\xi\rangle$ 
by a spin rotation matrix $D^{(S)}(\alpha,\beta,\gamma)$ corresponds to rotating 
the point configuration of $|\xi\rangle$ by $R(\alpha,\beta,\gamma)$ \cite{Bloch45}. 
Here $R(\alpha,\beta,\gamma)\in SO(3)$ is a rotation and 
$D^{(S)}(\alpha,\beta,\gamma)\in SU(2S+1)$ is its irreducible unitary $(2S+1)$- dimensional representation matrix. 
These rotations are parameterized by the Euler angles $\alpha,\beta,\gamma$, so we can be write  
\begin{equation}
\label{DS}
D^{(S)}(\alpha,\beta,\gamma)=\exp(-\rmi\alpha \mathbf{S}_z)\exp(-\rmi\beta \mathbf{S}_y)\exp(-\rmi\gamma \mathbf{S}_z),
\end{equation} 
where $\mathbf{S}_y$ and $\mathbf{S}_z$ are the spin operators of a spin-$S$ particle in $y$- and $z$-directions, respectively,  
and we have set $\hbar=1$. 

\section{Geometrical expression for $N$-qubit states using Majorana representation}
In this section, we present a way to express the pure states of an $N$-qubit system  
in terms of points on the Bloch sphere  
using the Majorana representation. 
As the Hilbert space of $N$ qubits is $2^N$ dimensional, every $N$-qubit state can be viewed as a state of a particle with spin $S=(2^N-1)/2$.  
We denote the basis of qubit $j$ by $\{|0\rangle_j,|1\rangle_j\}$.  Note that this basis is different from the 
one used earlier in the Majorana representation, written as $\{|+\rangle_j^{\textrm{M}},|-\rangle_j^{\textrm{M}}\}$.  
The basis vectors of an $N$-qubit system can be defined as 
\begin{equation}
|i\rangle_d\equiv|i_0\rangle_0\otimes|i_1\rangle_1\otimes\cdots\otimes |i_{N-1}\rangle_{N-1}, 
\end{equation}
where each $i_j\in \{0,1\}$ and we use the binary notation 
$i=\sum_{j=0}^{N-1} i_j 2^j$ to label the basis vectors. 
The subscript $d$ refers to decimal numbers.  
These $N$-qubit basis vectors can be expressed using the spin states of a spin-$S$ particle as   
\begin{equation}\label{correspondence}
|i\rangle_d=\sum_{j=0}^{2S} \Gamma_{ij}|S,j-S\rangle,\quad i=0,1,2,\ldots,2S,
\end{equation}
where $(\Gamma_{ij})\in U(2S+1)$ is an arbitrary unitary matrix. 
Let 
\begin{equation}
\phi = \sum_{i=0}^{2^N-1} C_i |i\rangle_d
\end{equation}
be some vector of the $N$-qubit system. Using (\ref{MaPo}) and (\ref{correspondence}) we see that the corresponding Majorana polynomial becomes 
\begin{equation}
\label{Majoqubit}
P_{Majo}(\phi;x)= \sum_{i,j=0}^{2^N-1}{2^N-1\choose j}^{1/2}   \Gamma_{ij} C_i\, x^j,
\end{equation}
where we have used the fact that $S=(2^N-1)/2$. We choose $(\Gamma_{ij})$ to be the identity matrix; hence
\begin{equation}
\label{Majoqubit2}
P_{Majo}(\phi;x)= \sum_{i=0}^{2^N-1}{2^N-1\choose i}^{1/2} C_i\, x^i.
\end{equation}
 Having fixed the mapping between the spin states and $N$-qubit states,  
we can now obtain a geometrical representation for any pure state $\phi$ by calculating the roots of $P(\phi;x)$ and using (\ref{angles}). 
As an example, we consider the case $N=2$. Then $S=\frac{3}{2}$ and the basis states are
\numparts
\begin{eqnarray}
\label{v00}
\fl |00\rangle =|0\rangle_d =|\frac{3}{2},-\frac{3}{2}\rangle = |---\rangle^{\textrm{M}}\\
\label{v10}
\fl |10\rangle =|1\rangle_d = |\frac{3}{2},-\frac{1}{2}\rangle = \frac{1}{\sqrt{3}}\left(|--+\rangle^{\textrm{M}}+|-+-\rangle^{\textrm{M}} +|+--\rangle^{\textrm{M}}\right)\\
\label{v01}
\fl |01\rangle=|2\rangle_d=|\frac{3}{2},\frac{1}{2}\rangle =\frac{1}{\sqrt{3}}\left(|++-\rangle^{\textrm{M}}+|+-+\rangle^{\textrm{M}}+|-++\rangle^{\textrm{M}}\right)\\
\label{v11}
\fl |11\rangle =|3\rangle_d=|\frac{3}{2},\frac{3}{2}\rangle = |+++\rangle^{\textrm{M}}.
\end{eqnarray}
\endnumparts
Here the first vector is a basis element of the two-qubit system, the second vector gives it  
in the basis labelled by decimal numbers, the third vector is the spin vector obtained from (\ref{correspondence}), 
and the rightmost vector gives its Majorana representation. We now consider two states defined as 
\begin{eqnarray}
\label{ent}
\phi_{\textrm{ent}} &=& \sqrt{3}|00\rangle+|10\rangle+|01\rangle+\sqrt{3}|11\rangle,\\
\label{sep}
\phi_{\textrm{sep}} &=& |00\rangle+|10\rangle+ |01\rangle+|11\rangle.
\end{eqnarray}  
It is easy to verify that $\phi_{\textrm{ent}}$ is an entangled state while $\phi_{\textrm{sep}}$ is separable. 
The point configurations do not depend on the normalization; hence these vectors have not been normalized. 
The Majorana polynomial (\ref{Majoqubit2}) of $\phi_{\textrm{ent}}$ reads 
\begin{equation}\label{2bitP}
P_{Majo}(\phi_{\textrm{ent}};x)=\sqrt{3}(1+x)(1+x^2).
\end{equation}
The roots are $x_0=-1$, $ x_1=\rmi$, and $x_2=-\rmi$; hence using (\ref{angles}) we see that the points are located at 
$(\frac{\pi}{2},\pi),(\frac{\pi}{2},\frac{\pi}{2})$, and $(\frac{\pi}{2},\frac{3\pi}{2})$, respectively;  see figure \ref{figure}(a). 
The points of $\phi_{\textrm{sep}}$ can be calculated similarly; the result is shown in figure \ref{figure}(d). 
The rotation of $\phi_{\textrm{ent}}$ and $\phi_{\textrm{sep}}$ by a spin-$\frac{3}{2}$ rotation matrix $D^{(3/2)}(\alpha,\beta,\gamma)$ 
moves the points as a rigid body. This is  illustrated by figures \ref{figure}(b) and \ref{figure}(e), 
where these states have been rotated by $D^{(3/2)}(0,\frac{\pi}{2},0)$. 
Spin-$\frac{3}{2}$ rotation can be written in terms of spin-$\frac{1}{2}$   
rotations as 
\begin{eqnarray}
\nonumber
\fl D^{(3/2)}(\alpha,\beta,\gamma)&=&D^{(1/2)}(\alpha,\beta,\gamma)^{\textrm{M}}\otimes D^{(1/2)}(\alpha,\beta,\gamma)^{\textrm{M}}\\
&&\otimes D^{(1/2)}(\alpha,\beta,\gamma)^{\textrm{M}}, 
\end{eqnarray}
where the right-hand side acts on the rightmost vectors of (\ref{v00})-(\ref{v11}), that is, 
on the spin-$\frac{1}{2}$ vectors of the Majorana representation. This is emphasized by the superscript M.  
However, when the states of an $N$-qubit system are considered, spin-$S$ rotations are not 
of particular interest. Instead of these one should concentrate on  single-qubit rotations, of which 
we now show two examples. We consider an identical simultaneous rotation of the two qubits, so the rotation operator reads   
\begin{equation}\label{2qubitrot}
D_{2}(\alpha,\beta,\gamma)\equiv D^{(1/2)}(\alpha,\beta,\gamma)\otimes D^{(1/2)}(\alpha,\beta,\gamma).  
\end{equation}
In figures \ref{figure}(c) and \ref{figure}(f), we show the states $\phi_{\textrm{ent}}$ and $\phi_{\textrm{sep}}$, respectively, 
after they have been rotated using $D_{2}(0,\frac{\pi}{2},0)$.  
Clearly now the points do not move as a rigid boby but their relative positions change. 

\section{Alternative geometrical representation}
In this section, we discuss another geometrical representation of $N$-qubit states, 
based on the polynomial presented in \cite{Makela09}.  
As before, $\phi$ is some $N$-qubit state expressed in a basis labelled by the decimal numbers 
\begin{equation}\label{somestate}
\phi=\sum_{i=0}^{2^N-1} \, C_i|i\rangle_d.  
\end{equation}
We associate this vector with the polynomial
\begin{equation}\label{CP}
P(\phi;x)\equiv \sum_{i=0}^{2^N-1}  C_i\,x^i.
\end{equation}
The only way to obtain this polynomial from the Majorana polynomial describing  $N$-qubit states, given by (\ref{Majoqubit}),  
 is to choose $\Gamma_{ij}=\delta_{ij}{2^N-1\choose j}^{-1/2}$.   
However, if $N>1$, this choice means that the matrix $(\Gamma_{ij})$ is not unitary and the left-hand side and 
right-hand side of equation (\ref{correspondence}) cannot both be normalized to one. 
Therefore, there is no unitary map between $N$-qubit states and spin-$S$ states which would  
produce $P(\phi;x)$ from the Majorana polynomial $P_{Majo}(\phi;x)$ for all states $\phi$. 

The benefits of (\ref{CP}) over the Majorana polynomial become clear when one studies the separability 
of pure $N$-qubit states \cite{Makela09} and the behaviour of the points in single-qubit rotations.  
In more detail, if $\phi_{\textrm{s}}$ is a separable $N$-qubit state
\begin{eqnarray}
\label{phis}
\phi_{\textrm{s}} 
&=& \bigotimes_{j=0}^{N-1} \left(a_j|0\rangle_j+b_j|1\rangle_j\right),
\end{eqnarray}
then the corresponding polynomial reads  
\begin{equation}
\label{Psep}
P(\phi_{\textrm{s}};x)=\prod_{j=0}^{N-1}(a_j+b_j x^{2^j}).
\end{equation}
A state $\phi$ is separable if and only if $P(\phi;x)$ can be written in this form \cite{Makela09}.  
 If $b_j\not=0$, (\ref{angles}) shows that the locations of the points determined by the qubit $j$ can be obtained from the equation 
\begin{equation}\label{thetajm}
\tan\frac{\theta_{jm}}{2}\exp(\rmi\varphi_{jm})=\left(-\frac{a_j}{b_j}\right)^{1/2^{j}}\!\!\!\!\!\!\exp\left(\rmi\frac{2\pi m}{2^j}\right),\quad m=0,1,\ldots,2^j-1. 
\end{equation}
If $b_j=0$, the points are at $(\theta_{jm}=\pi,\varphi_{jm}=0)$. Thus, for 
separable states the points always have a specific configuration. For example, 
if $N=2$, there are three points on the sphere. One of them can be anywhere while the remaining two are opposite to each other with respect 
to the $z$-axis. If $N=3$, seven points are needed, three of which 
are located as in the two-qubit case. The rest of the  points are at the vertices of a square orthogonal to the $z$-axis. 
If $N=4$, there are $15$ points on the Bloch sphere, seven of them are located as in the three-qubit case 
while the remaining ones are at the vertices of an octagon orthogonal to the $z$-axis. 
Proceeding in this way, one can obtain a general expression for the point configuration of an arbitrary  separable $N$-qubit pure state.

\begin{figure}[t]
\begin{center}
$\begin{array}{c@{\hspace{0.3cm}}c@{\hspace{0.3cm}}c@{\hspace{0.3cm}}c}
\multicolumn{1}{l}{\mbox{(a)}} &
	\multicolumn{1}{l}{\mbox{(b)}} & 
\multicolumn{1}{l}{\mbox{(c)}} \\ [0cm]
\includegraphics[scale=.35]{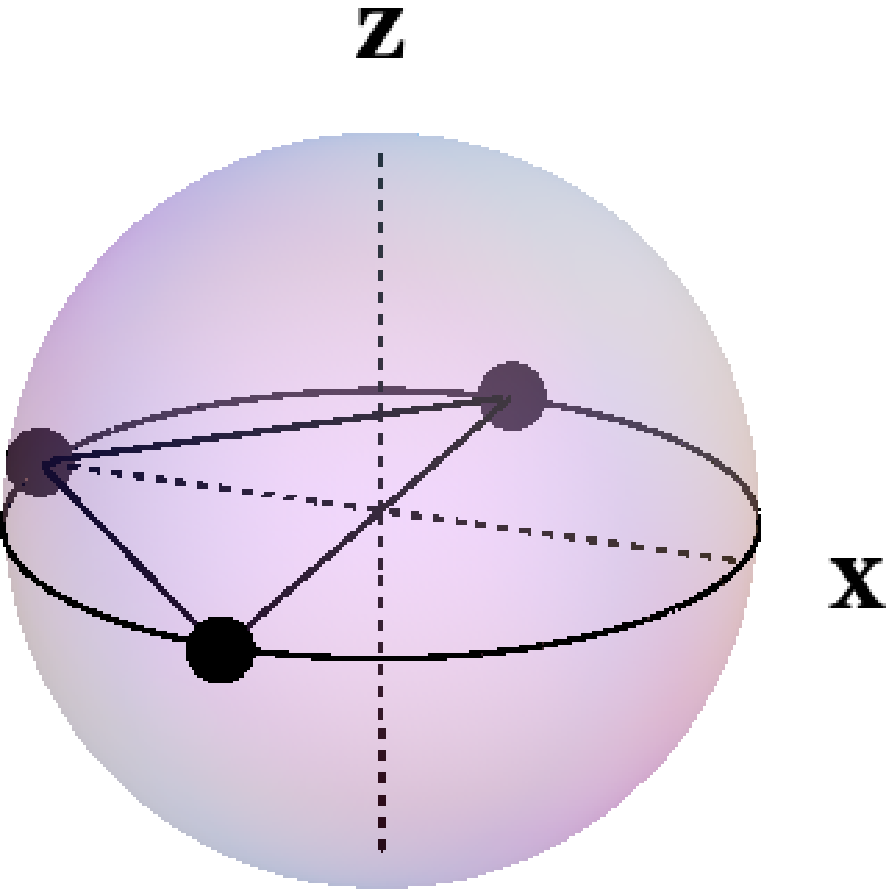} &
	\includegraphics[scale=.35]{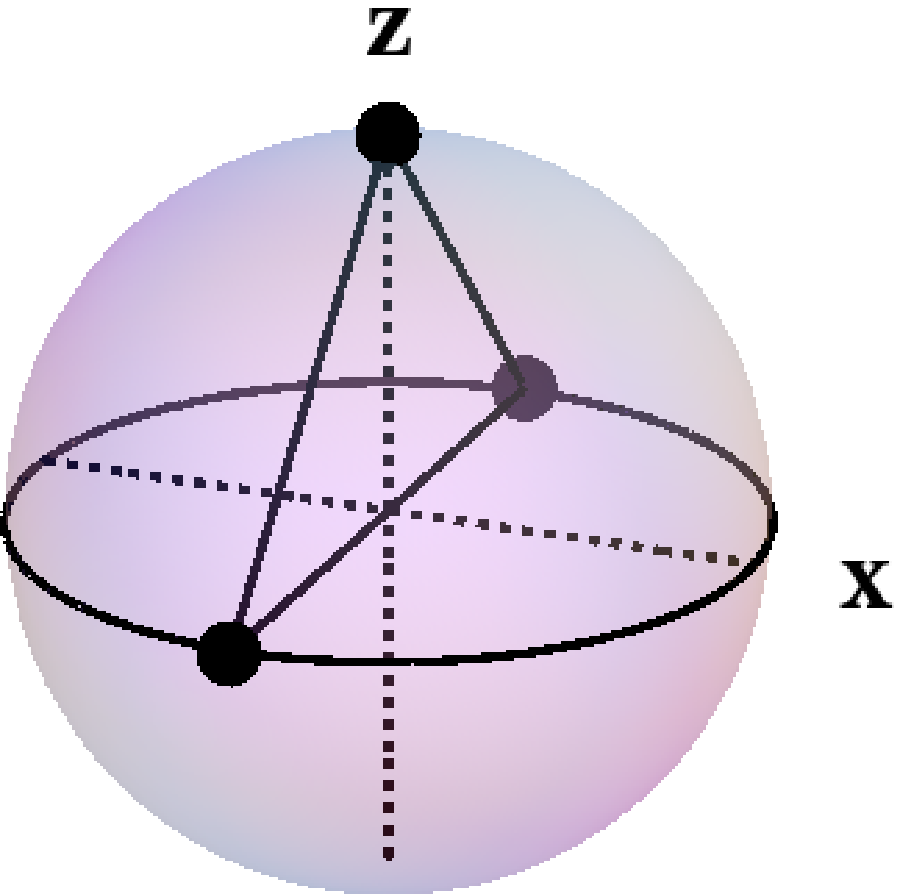} &
\includegraphics[scale=.35]{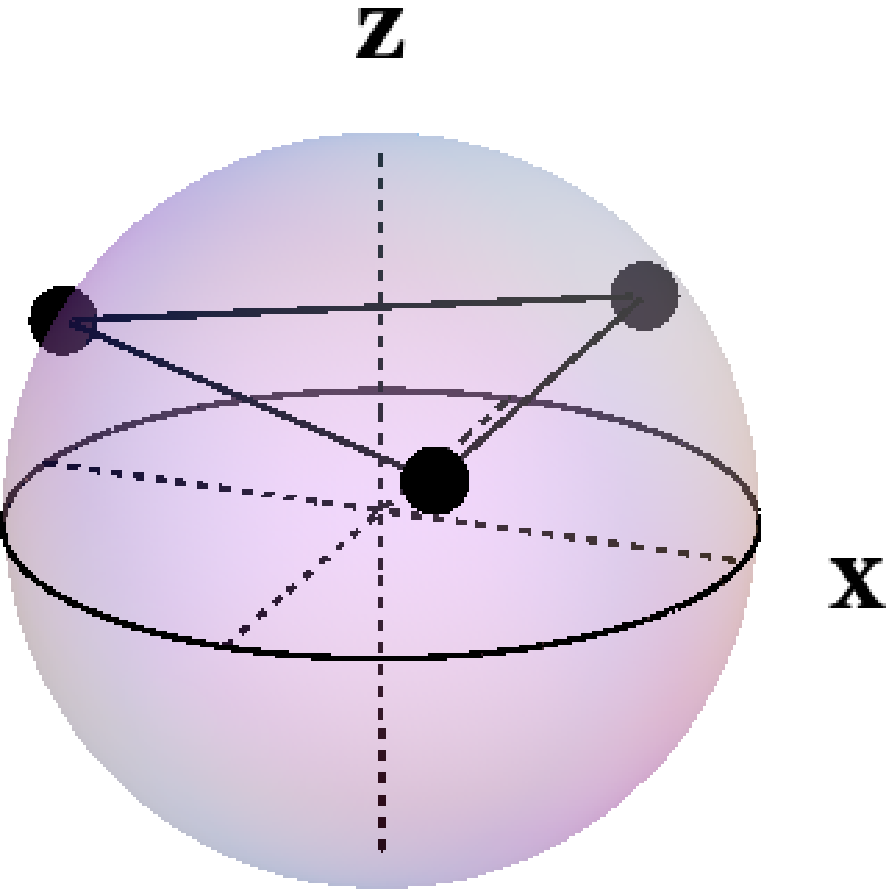} \\ [0cm]
\multicolumn{1}{l}{\mbox{(d)}} &
	\multicolumn{1}{l}{\mbox{(e)}} & 
\multicolumn{1}{l}{\mbox{(f)}} \\ [0cm]
\includegraphics[scale=.35]{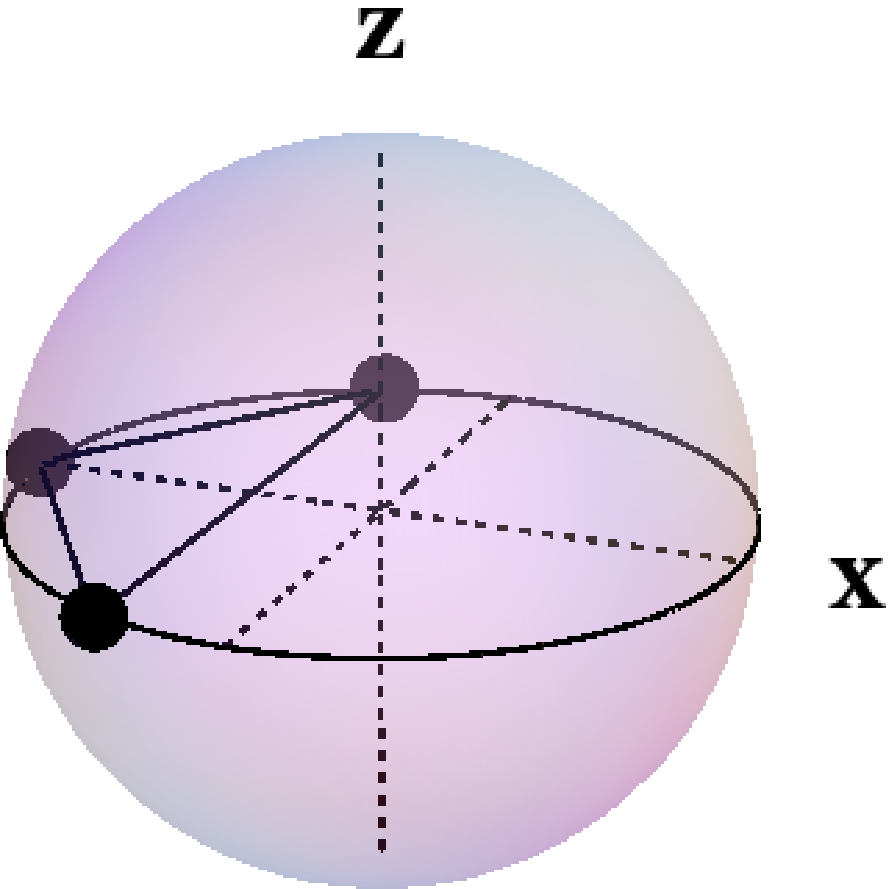} &
	\includegraphics[scale=.35]{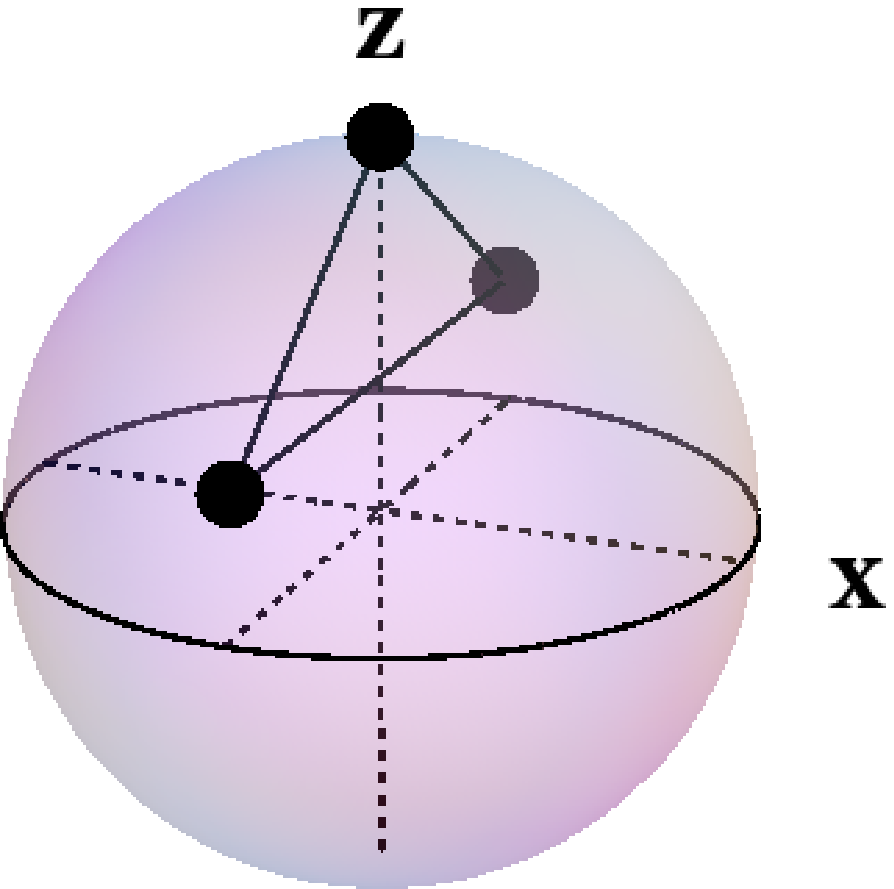} &
\includegraphics[scale=.35]{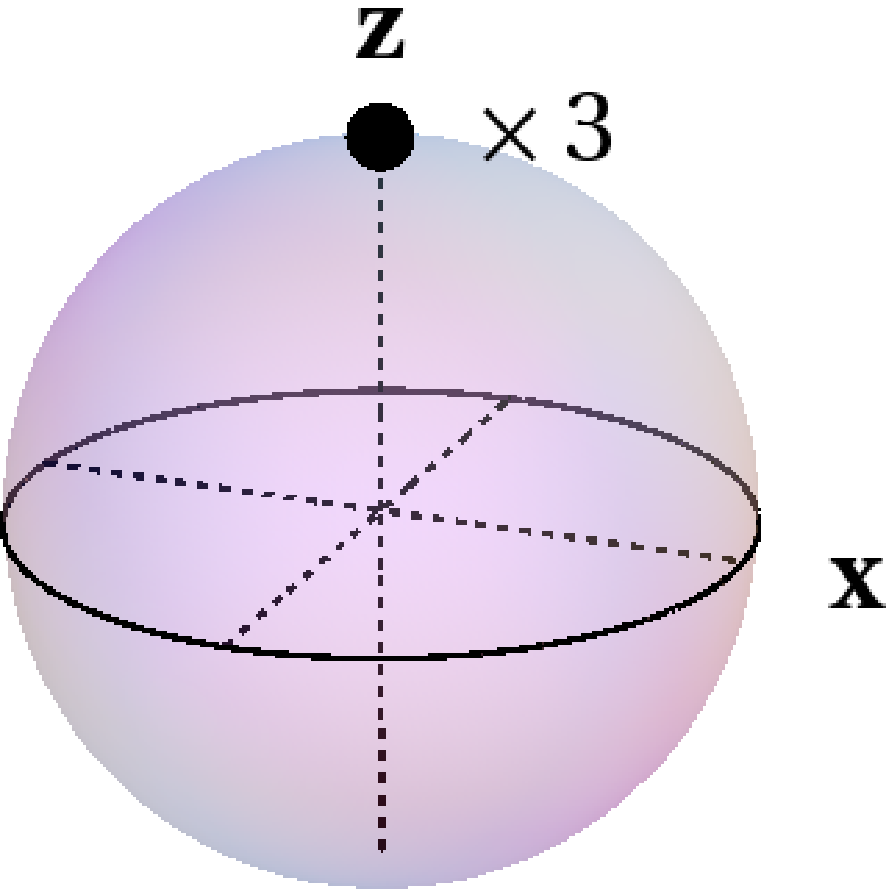} \\
\end{array}$
\end{center}
\caption{Majorana representation. 
(a) The entangled two-qubit state $\phi_{\textrm{ent}}$.
(b) $\phi_{\textrm{ent}}$ rotated through $\frac{\pi}{2}$ about the $y$-axis using spin-$\frac{3}{2}$ rotation. (c) $\phi_{\textrm{ent}}$ 
after the two qubits have been rotated through $\frac{\pi}{2}$ about the $y$-axis.  
(d)-(f) As in (a)-(c) but $\phi_{\textrm{ent}}$ replaced with $\phi_{\textrm{sep}}$. 
\label{figure}}
\end{figure}

\begin{figure}[t]
\begin{center}
$\begin{array}{c@{\hspace{0.3cm}}c@{\hspace{0.3cm}}c}
\multicolumn{1}{l}{\mbox{(a)}} &
	\multicolumn{1}{l}{\mbox{(b)}} & 
\multicolumn{1}{l}{\mbox{(c)}} \\ [0cm]
\includegraphics[scale=.35]{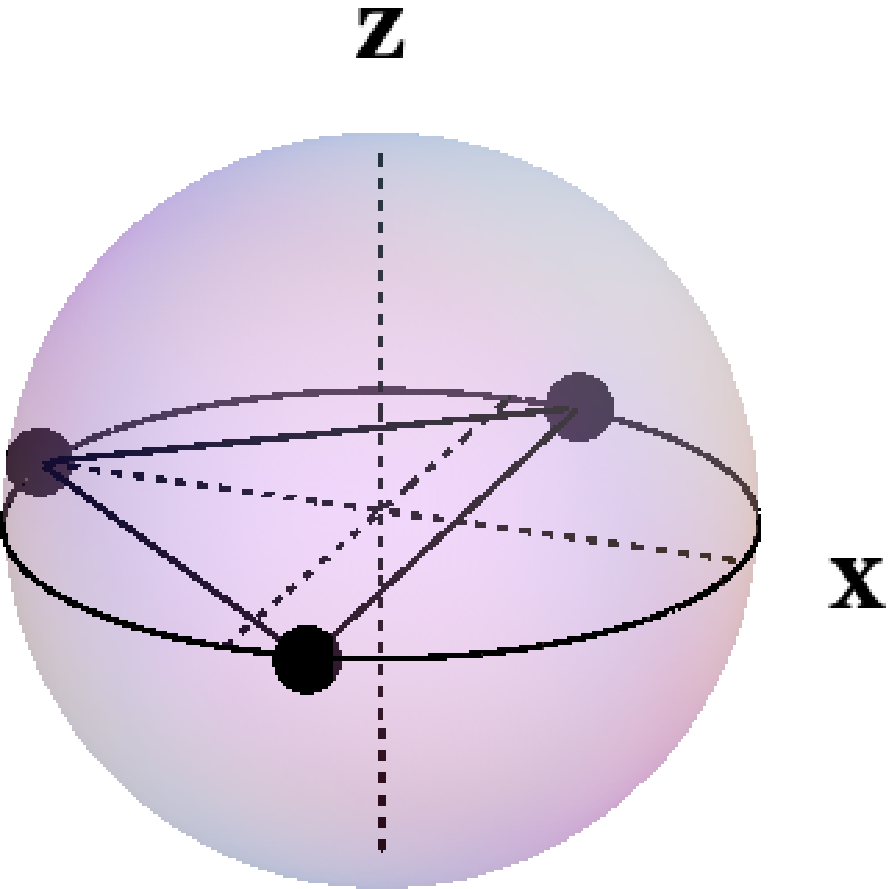} &
	\includegraphics[scale=.35]{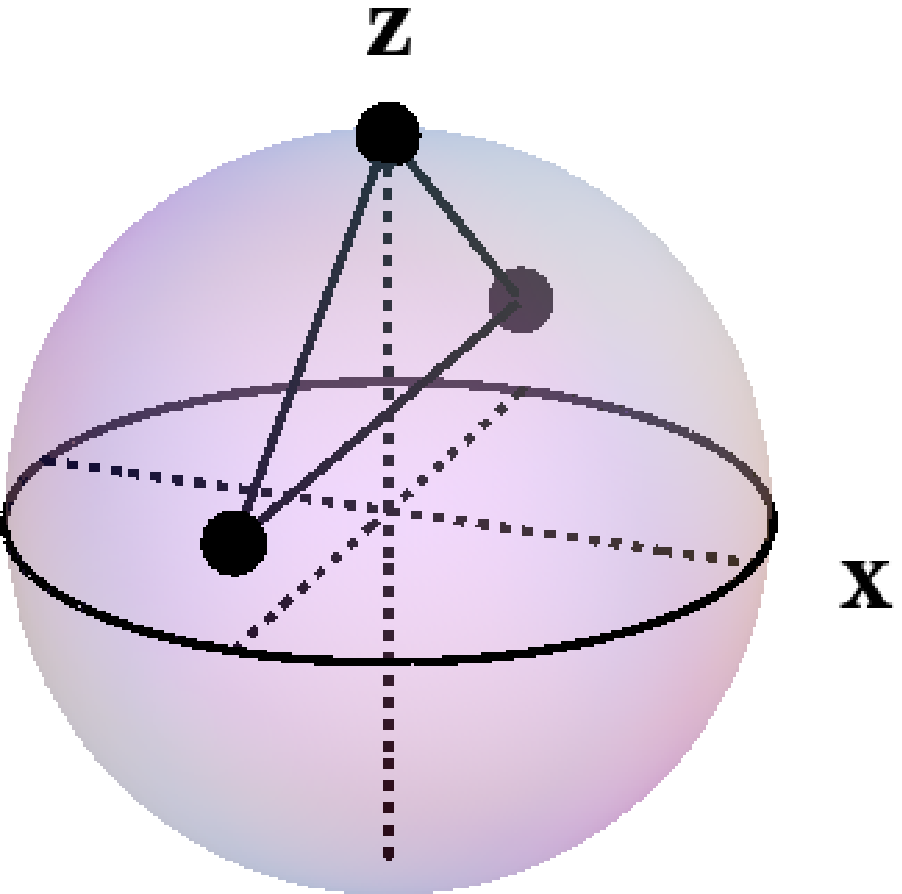} &
\includegraphics[scale=.35]{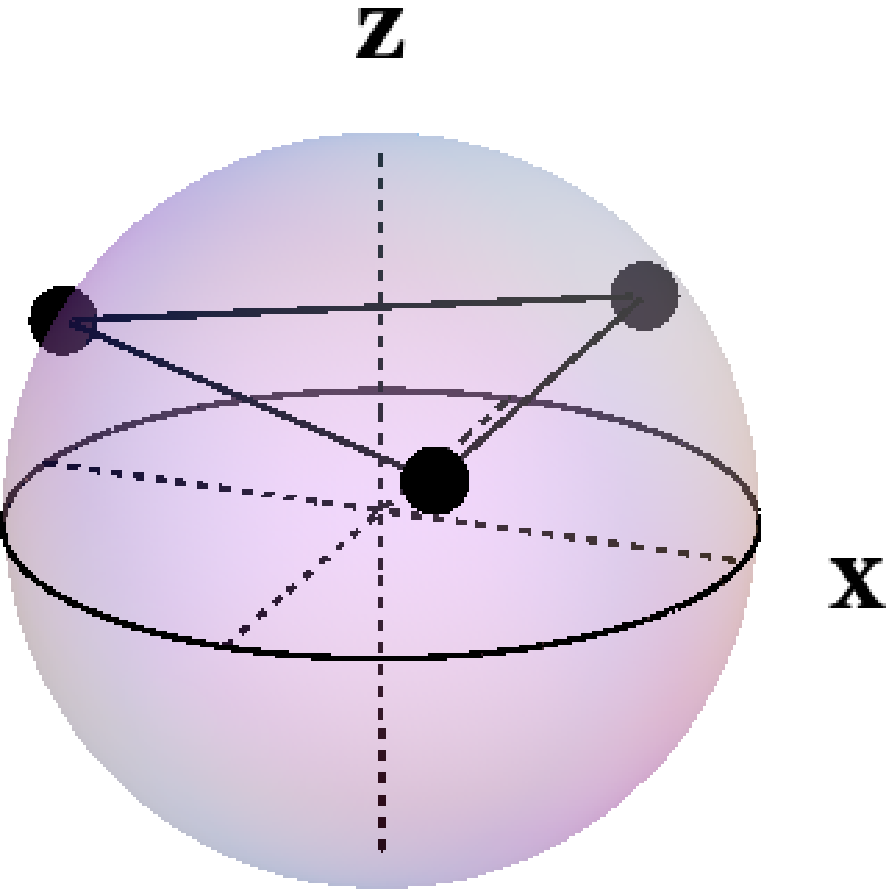} \\
\multicolumn{1}{l}{\mbox{(d)}} &
	\multicolumn{1}{l}{\mbox{(e)}} & 
\multicolumn{1}{l}{\mbox{(f)}} \\ [0cm]
\includegraphics[scale=.35]{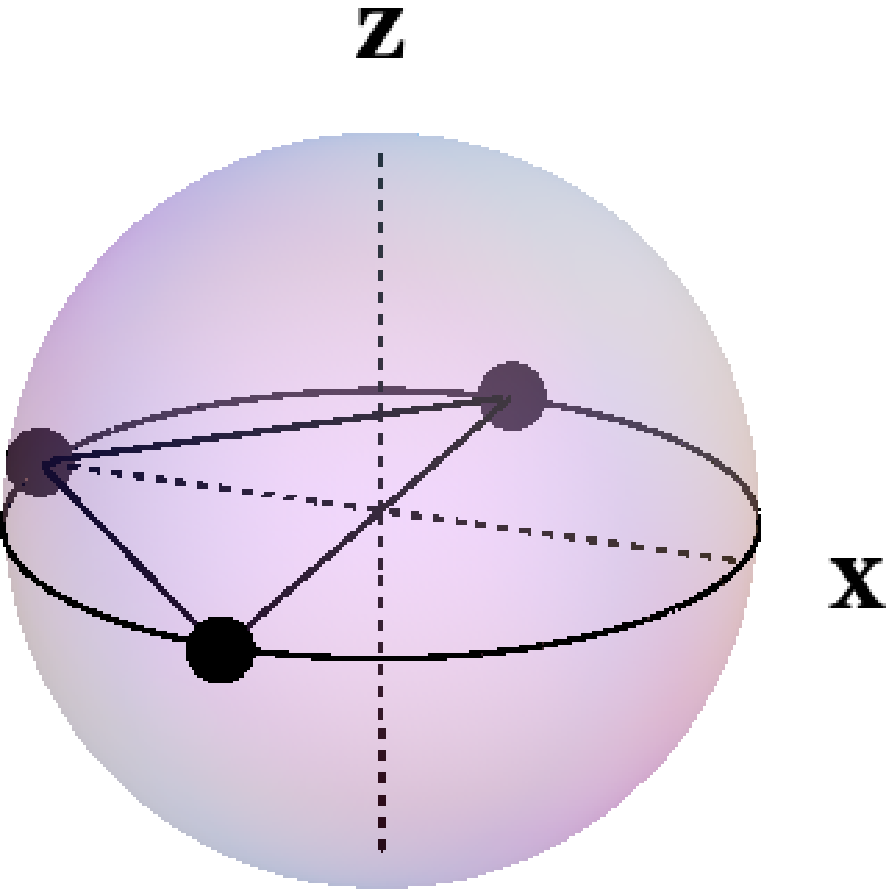} &
	\includegraphics[scale=.35]{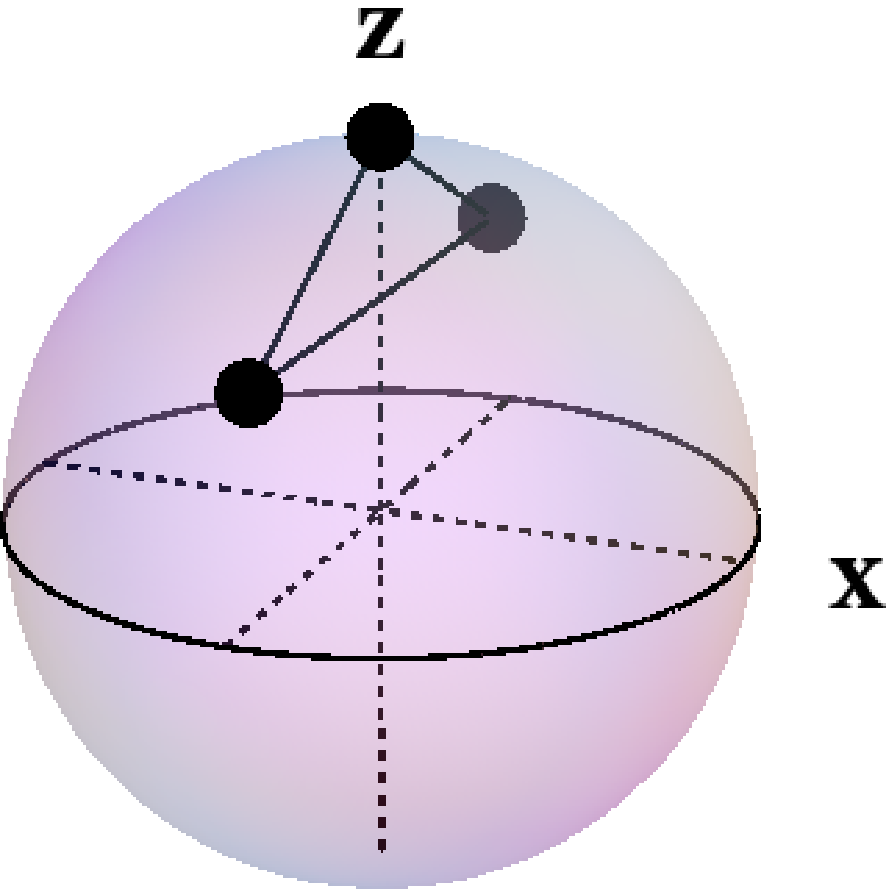} &
\includegraphics[scale=.35]{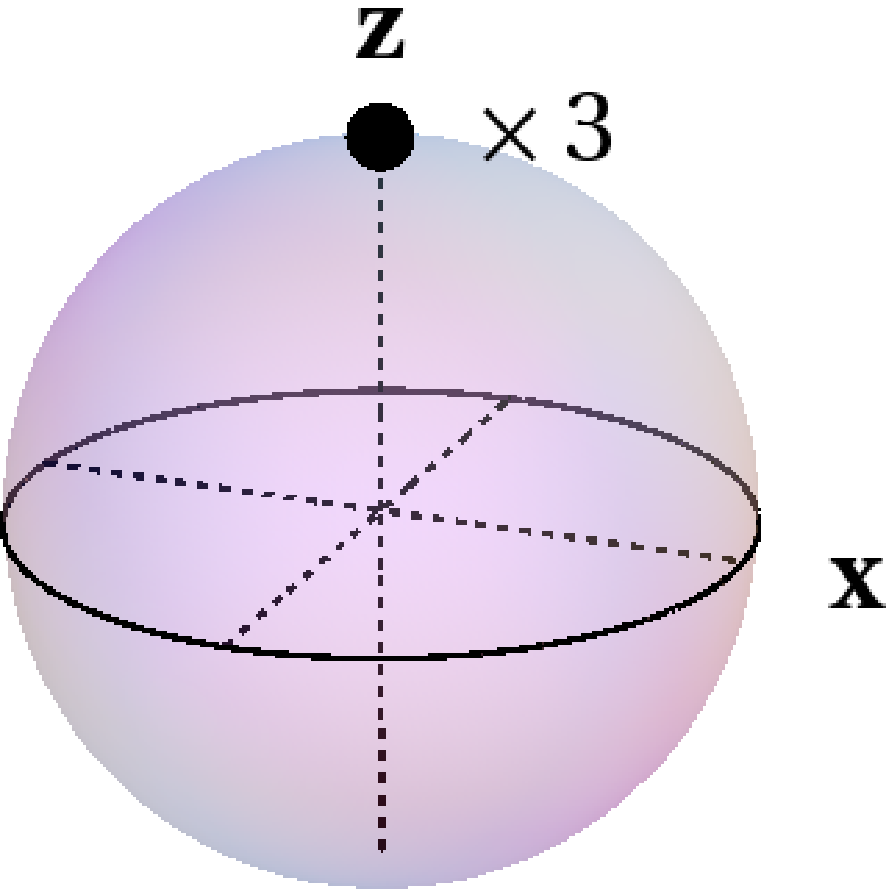} 
\end{array}$
\end{center}
\caption{Alternative representation. 
(a) The entangled two-qubit state $\phi_{\textrm{ent}}$. 
(b) $\phi_{\textrm{ent}}$ rotated through $\frac{\pi}{2}$ about the $y$-axis using spin-$\frac{3}{2}$ rotation. (c) $\phi_{\textrm{ent}}$ 
after the two qubits have been rotated through $\frac{\pi}{2}$ about the $y$-axis.    
(d)-(f) As in (a)-(c) but $\phi_{\textrm{ent}}$ replaced with $\phi_{\textrm{sep}}$.  
Two points located opposite to each other with respect to the $z$-axis shows that the state is separable. 
Therefore (b) and (d)-(f) are separable states, while (a) and (c) are entangled.   
\label{figure2}}
\end{figure}

The effect of $N$-qubit rotations on the point configurations of separable states  
can be straightforwardly obtained from (\ref{thetajm}). 
We denote by $D_{N}(\alpha,\beta,\gamma)$ the $N$-qubit generalization of $D_{2}(\alpha,\beta,\gamma)$:
\begin{equation}\label{DN}
D_{N}(\alpha,\beta,\gamma)\equiv \bigotimes_{j=0}^{N-1} D^{(1/2)}(\alpha,\beta,\gamma).
\end{equation}
A rotation of a separable state $\phi_{\textrm{s}}$ corresponds simply to 
applying $D^{(1/2)}(\alpha,\beta,\gamma)$ to all of the single-qubit states $\phi_j$. 
The components of the rotated states $\tilde{\phi}_j=\tilde{a}_j|0\rangle_j+\tilde{b}_j|1\rangle_j$ are 
\numparts
\begin{eqnarray}
\label{tildea}
\tilde{a}_j &=& a_j\cos\frac{\beta}{2}\, \exp\left(-\rmi\frac{\alpha+\gamma}{2}\right)-b_j\sin\frac{\beta}{2}\,
\exp\left(\rmi\frac{\gamma-\alpha}{2}\right)\\
\label{tildeb}
\tilde{b}_j &=& a_j\sin\frac{\beta}{2}\,\exp \left(-\rmi\frac{\gamma-\alpha}{2}\right)
+b_j\cos\frac{\beta}{2}\,\exp\left(\rmi\frac{\alpha+\gamma}{2}\right).
\end{eqnarray}
\endnumparts
which have been obtained using (\ref{DS}) and the spin matrices of a spin-$\frac{1}{2}$ particle. 
The locations of the points can then be obtained using (\ref{thetajm}). As  an example,  
we consider the state $\phi_{\textrm{sep}}$. The polynomial determined by this state is
\begin{equation}
P(\phi_{\textrm{sep}};x)=(1+x)(1+x^2),
\end{equation}
which is clearly of the form (\ref{Psep}), showing that $\phi_{\textrm{sep}}$ is a separable state. 
This polynomial equals the Majorana polynomial of $\phi_{\textrm{ent}}$; hence the point configuration is the same, 
as can be seen by comparing figures \ref{figure}(a) and \ref{figure2}(d).  
A two-qubit rotation of 
$\phi_{\textrm{sep}}$ through $\frac{\pi}{2}$ about the $y$-axis corresponds to $\alpha=\gamma=0,\beta=\frac{\pi}{2}$. 
In the initial state $a_0=b_0=a_1=b_1=1$, hence, using (\ref{tildea}), (\ref{tildeb}) we see that in the rotated state 
$\tilde{a}_0=\tilde{a}_1=0$, $\tilde{b}_0=\tilde{b}_1=1$, so all the points are at the north pole, see 
figure \ref{figure2}(f). 
In deriving (\ref{tildea}), (\ref{tildeb}) it was assumed that all qubits are rotated by the same single-qubit rotation.  
However, it is easy to generalize this treatment to the case 
where these rotations are not identical. Then the angles in (\ref{tildea}), (\ref{tildeb}) have to 
be replaced by qubit-dependent angles. Like before, the locations of points can be obtained from (\ref{thetajm}). 
One should note that this discussion applies only to separable states;  
it is not known at the moment how the points corresponding to an entangled  
$N$-qubit state move under these type of rotations.

Unlike in the case of the Majorana representation, now spin-$\frac{3}{2}$ rotations change the relative locations of the points. 
For example, a spin-$\frac{3}{2}$ rotation of $\phi_{\textrm{sep}}$ by $D^{3/2}(0,\frac{\pi}{2},0)$ moves one of the points to the north pole and 
shifts the other two points upwards by the same amount, keeping the state separable, see figure \ref{figure2}(e). 
The point configuration corresponding to $\phi_{\textrm{ent}}$ is presented figure \ref{figure2}(a). 
As in the case of $\phi_{\textrm{sep}}$, now a spin rotation by a spin-$\frac{3}{2}$ rotation matrix 
$D^{(3/2)}(0,\frac{\pi}{2},0)$ does not correspond to a rigid body rotation of the points, see figure \ref{figure2}(b).   
Interestingly, the final state is separable; hence this rotation renders the entangled state $\phi_{\textrm{ent}}$ separable.

\section{Conclusions}
In this paper, we have discussed two different geometrical representations of pure $N$-qubit states.  
We have first presented a way to write spin-$S$ states in terms of symmetrized states of $2S$ 
spin-$\frac{1}{2}$ systems. We have then shown how this method, known as the Majorana representation, 
can be used to characterize the states of an $N$-qubit system and how these states can be given a 
geometrical representation as points on the Bloch sphere. 
We have compared this geometrical representation with another one, 
defined recently by the present authors \cite{Makela09}. 
The Majorana representation is well suited for the description of 
spin-$S$ particles as a spin rotation by a spin-$S$ rotation matrix corresponds to a 
rigid body rotation of the point configuration of the state.  
However, in the case of $N$-qubit systems the interesting operations are not spin-$S$ rotations but single qubit ones.  
We have shown through simple examples that in the latter rotations the points of the Majorana representation move in a complex way. 
Furthermore, using the Majorana representation it is not easy to see the separability of states from their point configurations.   
In contrast to this, in the alternative representation separability can be detected, at least in pricinple, from the   
point configurations. Additionally, the behaviour of points in single-qubit rotations can be straightforwardly calculated if the 
rotated state is separable. In conclusion, the Majorana representation is useful when spin-$S$ particles are studied, 
while the alternative representation is preferable when the states of an $N$-qubit system are discussed.  
Besides helping to visualize $N$-qubit states and the way they transform 
in rotations and other operations, the latter representation may also help to identify some special  $N$-qubit states, 
like the Majorana representation did in the context of spinor Bose-Einstein condensates \cite{Makela07}.
\ack
The authors are grateful to V I Man'ko and M A Man'ko for helpful discussions. 
AM acknowledges partial support by MIUR project II04C0E3F3 
Collaborazioni Interuniversitarie ed Internazionali Tipologia C. 
HM thanks E Kyoseva, B W  Shore, and N V  Vitanov for discussions  
and EC projects CAMEL and EMALI for financial support.

\end{document}